# Atomically precise lateral modulation of a two-dimensional electron liquid in anatase TiO$_2$ thin films

Z. Wang[1,2], Z. Zhong[3], S. McKeown Walker[2], Z. Ristic[4], J.-Z. Ma[5], F. Y. Bruno[2], S. Riccò[2], G. Sangiovanni[3], G. Eres[6], N. C. Plumb[1], L. Patthey[1,7], M. Shi[1], J. Mesot[1,4,8], F. Baumberger[2,1] and M. Radovic[1,7]

[1]*Swiss Light Source, Paul Scherrer Institut, CH-5232 Villigen PSI, Switzerland*

[2]*Department of Quantum Matter Physics, University of Geneva, 24 Quai Ernest-Ansermet, 1211 Geneva 4, Switzerland*

[3]*Institut für Theoretische Physik und Astrophysik, Universität Würzburg, Am Hubland, Germany*

[4]*Institute of Condensed Matter Physics, École Polytechnique Fédérale de Lausanne (EPFL), CH-1015 Lausanne, Switzerland*

[5]*Beijing National Laboratory for Condensed Matter Physics, and Institute of Physics, Chinese Academy of Sciences, Beijing 100190, China*

[6]*Materials Science and Technology Division, Oak Ridge National Laboratory, Oak Ridge, Tennessee 37831, United States*

[7]*SwissFEL, Paul Scherrer Institut, CH-5232 Villigen PSI, Switzerland*

[8]*Laboratory for Solid State Physics, ETH Zürich, CH-8093 Zürich, Switzerland*

**Abstract**

Engineering the electronic band structure of two-dimensional electron liquids (2DELs) confined at the surface or interface of transition metal oxides is key to unlocking their full potential. Here we describe a new approach to tailoring the electronic structure of an oxide surface 2DEL demonstrating the lateral modulation of electronic states with atomic scale precision on an unprecedented length scale comparable to the Fermi wavelength. To this end, we use pulsed laser deposition to grow anatase $TiO_2$ films terminated by a (1 × 4) in-plane surface reconstruction. Employing photo-stimulated chemical surface doping we induce 2DELs with tunable carrier densities that are confined within a few $TiO_2$ layers below the surface. Subsequent *in-situ* angle resolved photoemission experiments demonstrate that the (1 × 4) surface reconstruction provides a periodic lateral perturbation of the electron liquid. This causes strong backfolding of the electronic bands, opening of unidirectional gaps and a saddle point singularity in the density of states near the chemical potential.

**Main text**

Two dimensional electron liquids (2DELs) confined at the surface[1-3] or interface[4] of transition metal oxides exhibit fascinating properties such as superconductivity[5], negative compressibility[6] and large thermoelectric efficiency[7] that can be controlled by tuning the carrier density and are of interest for applications beyond those of current semiconductor technology. Experimentally, carrier density control was achieved by field-effect gating[8], interface engineering[9], or surface doping[1,10-14]. Achieving comparable control over the band structure of oxide 2DELs and thin films would open new opportunities for the emerging field of oxide electronics. To date such efforts have largely focused on varying the surface orientation and exploiting interfacial lattice mismatch. For instance, in $SrTiO_3$ - based 2DELs it was demonstrated that changing the crystalline orientation strongly modifies the orbital polarization, and affects spin-orbit coupling and superconductivity[3,13,15-17]. Lateral confinement was achieved in a top-down approach and exploited in nano-electronic devices[10,11]. On the other hand, bottom-up microscopic electronic structure engineering based on intrinsic atomic scale lateral reconstructions, which are ubiquitous[18,19] in oxides, has received little attention.

Here we apply this concept, originally developed for semiconductor and elemental metal surfaces[20,21], to electronic structure engineering of an oxide 2DEL. Investigating a surface 2DEL on anatase $TiO_2$ (001) thin films we show that a long range ordered surface reconstruction[22,23] causes a periodic perturbation of the electronic structure. Tuning the 2DEL carrier density such that its Fermi wave vector matches the wave vector of the lateral superstructure, we induce a unidirectional band gap and a van Hove singularity near the chemical potential. These findings open a new pathway for tailoring the electronic properties of oxide 2DELs.

Oxide 2DELs are observed at suitably chosen interfaces [4] but can also be created at bare oxide surfaces where they are accessible to high-resolution angle-resolved photoemission (ARPES)[1-3,12-14,24-29]. Such ARPES studies on $SrTiO_3$ and anatase $TiO_2$ revealed strong many-body effects[12,24,25] and high carrier densities of $10^{13} - 10^{14}$ cm$^{-2}$ [2,14], approaching the 0.5 electrons/unit cell predicted for the ideal (001) $LaAlO_3/SrTiO_3$ interface[30]. The short electronic length scales associated with such carrier densities pose a serious challenge for electronic structure engineering by lateral patterning because a large response is only expected if the confinement length approaches the Fermi wavelength $\lambda_F$. For a single isotropic band $\lambda_F = (2\pi/n_{2D})^{1/2} = 2.5 - 8$ nm for the aforementioned densities. Defining features on these length scales with a top-down approach runs into fundamental physical limits. Searching for an alternative route to nanoscale patterning, we chose to investigate the electronic band structure of a highly itinerant 2DEL in anatase $TiO_2$ thin films terminated by a periodic surface reconstruction. The weakly anisotropic bulk crystal structure and corresponding Brillouin zone of anatase $TiO_2$ are shown in Figures 1a-1b. Anatase $TiO_2$ is particularly suitable for experiments probing the effects of lateral surface patterning on the properties of oxide 2DELs since it is amenable to electron doping by cation substitution[31] or the creation of oxygen vacancies[25,26]. Moreover, as shown in Figure 1c, its conduction band minimum at the $\Gamma$ point has pure $d_{xy}$ orbital character, which strongly enhances the confinement in a band-bending potential[24,32].

Experimentally, we induce a surface band-bending potential that quantum-confines electrons along the $z$ direction by exposing the anatase $TiO_2$ thin films to synchrotron UV light at a temperature < 20 K (see Figure 2a and Supplementary Information Section II for more details). The surface electron doping is caused by formation of surface oxygen vacancies through photon-induced desorption of $O^+$ ions. This process has been recognized early by Knotek and Feibelman for $TiO_2$[33], and was recently used to create 2DELs on other oxides including $SrTiO_3$, $KTaO_3$ [1-3,34-36]. By controlling the irradiation time, we can tune the carrier density. As the irradiation time increases, the carrier density and corresponding Fermi wave vector increases and eventually saturates. In Figure 2b and 2c we show the photoemission intensity map of the saturated states, and the irradiation time dependent momentum distribution curves (MDC) taken at $E_F$. Note that the photoemission intensity of the metallic states is suppressed when measured around the $\Gamma_{00}$ point due to the photoemission matrix effect, thus the ARPES data shown below were measured around the $\Gamma_{10}$ point (see Supplementary Information Section IV for more details). In Figures 2d-2f, we characterize the resulting saturated 2DEL. The first clear signature of quantum confinement is the presence of three peaks in the energy distribution curve (EDC) taken at the $\Gamma_{10}$ point in an energy range where the bulk band structure shows only a single state. We associate the experimentally observed peaks with the $n = 1,2,3$ quantum well states of the $d_{xy}$ bulk conduction band in line with a previous study that observed two quantum well states at slightly lower doping[26]. From Lorentzian fits, we obtain occupied band widths of ~ 170, 50, and 10 meV, respectively, which are reproduced by our surface band structure calculation shown in Figure 4 (see also Figure S5 in Supplementary Information).

The 3D ARPES intensity maps in the $k_x$-$k_y$ surface plane and the $k_x$-$k_z$ plane perpendicular to the surface are presented in Figures 2d and 2e. We find dispersive metallic states with concentric circular Fermi surfaces in the $k_x$-$k_y$ plane, consistent with a $d_{xy}$ orbital character. The lack of dispersion along the surface normal $k_z$ (Figure 2e and Supplementary Information Figure S6) confirms that the states are highly two-dimensional as expected for quantum-confined subbands in a 2DEL, in sharp contrast with the 3D polaronic state reported in Ref. 25 for a lower carrier density. We further confirmed the $d_{xy}$ orbital character of the 2DEL directly by light polarization dependent measurements (see Supplementary Information Figure S4).

We note that the spectra show pronounced many-body effects in the form of a broad replica band below the band bottom and an abrupt change of dispersion ("kink") at E – $E_F$ ≈ -70 meV (see Figures 2b and 3d). These observations are broadly consistent with earlier measurements on bulk-like states in anatase $TiO_2$ [25] and will not be discussed in details here. Instead we focus our attention to the lateral confinement effect on the 2DEL at the surface of anatase $TiO_2$. The mapping of the 2DEL electronic structure over an extended $k$-space area shown in Figure 3 demonstrates the pronounced effect of the surface superstructure on the subband dispersion. Umklapps of the main band following the periodicity G = 0.5 π/a of the (1 × 4) surface reconstruction (here a = 3.9 Å is the lattice constant of the $SrTiO_3$ substrate) are evident in the Fermi surface map in Figures 3a-3b. In order to maximize the effect of the surface superstructure on the dispersion we tuned the Fermi wave vector $k_F$ of the first subband to match G = 0.5 π/a by optimizing the surface doping through controlling the irradiation dose (see Supplementary Information Section III for more details). For this band filling, the main and umklapp band of the $n$ = 1 quantum well state cross at the chemical potential (see Figure 3d) where they interact opening a unidirectional band gap which breaks the four-fold rotational symmetry of the bulk band structure.

The quantification of this gap is non-trivial since our photoemission experiments average over two perpendicular domains of the surface reconstruction, which are evident in both low energy electron diffraction (LEED) pattern and Fermi surface map (Figures 3a-3b). This leads to a superposition of gapped and ungapped spectral weight at the first Brillouin zone boundary of the super-structure at $k_{x,y}$ = 0.25 π/a as sketched in Figure 3c. We therefore focus on the Brillouin zone boundary at $k_{x,y}$ = 0.75 π/a where the spectral weight is dominated by domains with a single orientation.

Comparing EDCs extracted at these points shows a clear shift of the leading edge indicative of a super-lattice band gap (Figure 3e). In Figure 3f we quantify this gap from fits to a Lorentzian quasiparticle peak multiplied by a Fermi function. The experimental resolution is taken into account by a Gaussian convolution. This analysis shows a peak position of -20 meV at $k_x = 0.75\ \pi/a$, providing a lower limit of the band gap. In order to estimate the full gap, which will extend into the unoccupied states, we recall that the band filling is chosen such that $k_F \approx 0.25\ \pi/a$. This suggests approximate particle hole symmetry and thus a full band gap size of $\approx 40$ meV, well above $k_B T$ at room temperature.

The electronic structure calculations shown in Figure 4 provide further insight into the properties of this nano-patterned 2DEL. Our starting point is a density functional calculation of the (1 × 4) reconstructed surface using a slab geometry based on the structural model of Lazzeri and Selloni (Figure 4a)[22]. The layer-resolved Ti $3d$ density of states from this calculation shows a tail contributed by subsurface Ti atoms below the bulk conduction band minimum (Figure 4c). This indicates that the 2DEL resides at the subsurface $TiO_2$ layers, which are capped by the topmost (1 × 4) reconstruction layer.

In order to overcome the restrictions in unit cell size of full *ab-initio* calculations, we study the combined effect of quantum confinement and the lateral potential modulation on the 2DEL electronic structure with a realistic tight binding model [24,37]. In a first step, we estimated the perpendicular confinement potential due to surface band bending from a self-consistent solution of coupled Poisson-Schrödinger equations using an *ab-initio* tight-binding Hamiltonian for a supercell comprising 30 unit cells normal to the surface. The conduction band minimum in the surface layer is treated as a free parameter in this calculation and is chosen such as to reproduce the experimentally observed binding energy of the $n = 1$ quantum well state. Subsequently, we model the lateral superstructure with an onsite potential term included in the first layer of a 4 × 30 supercell. We estimated this potential term from the site resolved Ti 3s semi core level spectra calculated within DFT (Figure 4b) which show a variation of $\approx 120$ meV in the local electrostatic environment of the surface Ti sites.

The spectral function calculated in this model reproduces the key experimental observations. We first note that all three subband energies are in good agreement with experiment, indicating that the calculation converges to a realistic confinement potential along the surface normal. Clearly, it also reproduces the backfolded bands and accounts for their weak spectral weight (Figure 4d), which remains nearly constant throughout the second Brillouin zone of the superstructure. The calculated band gap in the dominant first subband is 56 meV, in fair agreement with the experimental estimate of ≈ 40 meV. For the higher subbands, we find much smaller band gaps of < 4 meV. This can be traced back to the rapidly decreasing amplitude of the wave function in the surface layer, which is affected by the lateral potential modulation. Indeed, our calculations show that the $n = 1$ subband has around 90% of the total charge in the first unit cell of the model (which contains two $TiO_2$ planes), whereas this value decreases to 10% and 4% for the second and third subband. This emphasizes the importance of the $d_{xy}$ orbital character of the 2DEL in anatase $TiO_2$. Since $d_{xy}$ states have a heavy effective mass for motion perpendicular to the surface, their confinement energies are small. Hence, the wave function of the $n = 1$ $d_{xy}$ subband is highly confined near the bottom of the potential well at the surface where the periodic perturbation from the superstructure is maximal. The out-of-plane $d_{xz,yz}$ states, which are important in $SrTiO_3$ [38], on the other hand have generally more spatially extended subband wave functions and would be less susceptible to a lateral confinement potential in the surface layer.

For 2D states, as they are observed here, super-lattice band gaps cause extended $k$-space areas around the reduced Brillouin zone boundary in which the transport effective mass m* = $\hbar^2(d^2E/dk^2)^{-1}$ is negative along one direction and positive along the perpendicular direction. Such saddle-points in the dispersion cause a divergent density of states (DOS) at the gap edge, which we observe at -20 meV for the carrier density investigated in our experiments. In Figure 4f we illustrate this effect by plotting the DOS of the first unit cell along the surface normal from our supercell calculations. This clearly confirms the presence of a singularity at the lower gap edge. Using a field effect device to tune such a saddle-point singularity through the chemical potential might open a new route towards controlling many-body states in oxide 2DEGs. Moreover, using single-domain thin films and nanoscale probe systems it should be possible to study the anisotropy of transport properties in such systems [39].

Our work reveals that the surface reconstruction of anatase TiO$_2$ thin films can be used as an atomic scale mask for the lateral modulation of electronic states in an oxide 2DEL. More generally, the bottom-up route to nano-patterning presented here might be utilized to selectively tune the lateral modulation of subbands with different orbital character as they are found at the LaAlO$_3$/SrTiO$_3$ interface[4] and the SrTiO$_3$ (001) surface[1,2,12] or to tailor shape resonances in superconducting 2DELs [40-42]. Combining the unique properties of 2DELs at complex oxide surfaces and interfaces with suitable engineering of surface structures thus provides a new platform for creating novel atomic scale functionality.

**Methods**

**Sample preparation.** The anatase TiO$_2$ (001) thin films were grown by PLD on TiO$_2$ terminated, 0.5 wt% Nb-doped SrTiO$_3$ (001) substrates supplied by CrysTec, GmbH[43]. The substrate was heated resistively to 750 °C by passing a direct current through it while monitoring the temperature by an infrared pyrometer. The oxygen partial pressure during thin film growth was $1 \times 10^{-5}$ mbar. The film growth was monitored by reflection high-energy electron diffraction and the film thickness is estimated as ~10 nm from the growth rate and deposition time. The surface structure of the thin films was verified by LEED immediately after transferring in the ARPES chamber.

**ARPES**. Following film growth, the samples were cooled to room temperature at the same O$_2$ pressure, and transferred *in-situ* to the ARPES chamber. ARPES measurements were performed at the SIS beamline of the Swiss Light Source in a temperature range of $T$ = 16 - 150 K, and photon energy range of $hv$ = 30 - 110 eV. The energy and momentum resolutions were ~ 20 meV and 0.2°, respectively. To estimate the $k_z$ dispersion from photon-energy-dependent measurements, we employed a free electron final state model with an inner potential of 13 eV[25,26].

**Calculations.** DFT calculations of the bulk and surface electronic structure were performed with the Wien2k and VASP packages, respectively [44,45]. The tight-binding calculations of the surface electronic structure use a Hamiltonian derived from transfer integrals obtained from a downfolding of a bulk DFT calculation onto maximally localized Wannier functions. The quasiparticle mass enhancement observed experimentally is taken into account by a simple renormalization of all transfer integrals by a factor of 0.6. The band bending confinement potential is included as an on-site potential term in a supercell containing 30 unit cells along the surface normal. This Hamiltonian is solved self-consistently with the Poisson equation using a field dependent dielectric constant. The lateral superstructure is included in a second step without repeating the self-consistent cycle by introducing the periodic potential in Figure 4a in the surface layer of a 4 × 30 supercell.

**Acknowledgments** We thank Christophe Berthod, Ulrike Diebold, Martin Setvin and Anna Tamai for helpful discussions. Experiments were conducted at the Surface/Interface Spectroscopy (SIS) beamline of the Swiss Light Source within the Paul Scherrer Institut in Villigen, Switzerland. SMW, FYB and FB were supported by the Swiss National Science Foundation (Grant No. 200021-146995). The contribution by GE was sponsored by the Materials Sciences and Engineering Division, Basic Energy Sciences, Office of Science, U.S. Department of Energy.

**Figure captions**

Figure 1. **The crystal and electronic structure of anatase $TiO_2$**. **a**, Crystal structure of anatase $TiO_2$ consisting of chains of vertex-linked distorted $TiO_6$ octahedra that share alternating edges along the *c* axis. Each octahedron also shares corners with four neighbors in the plane. **b**, Bulk and surface Brillouin zone. **c**, DFT calculated bulk band structure with color coded orbital character.

Figure2. **Creation of 2DELs at the surface of anatase $TiO_2$ (001) thin films**. **a**, Schematic illustration of synchrotron UV light irradiation induced a 2DEL at anatase $TiO_2$ surface. **b**, ARPES energy-momentum intensity map of a 2DEL with saturated band width measured around the $\Gamma_{00}$ point. **c**, Evolution of the momentum-distribution curves taken at EF as a function of irradiation time. **d**, 3D ARPES intensity maps in the $k_x$-$k_y$ surface plane, exhibiting two concentric circular electron pockets. **e**, 3D ARPES intensity maps in the $k_x$-$k_z$ plane perpendicular to the surface, showing no sign of dispersion along $k_z$. **f**, EDC at $k_{\bar{X}} = 2\pi/a$ labeled by the white dashed line in **e**. The spectrum is fitted by three Lorentzians (green), representing the subbands with $n$ = 1, 2 and 3. The ARPES data in **d**-**f** were taken around the $\Gamma_{10}$ point.

Figure 3. **The effect of the surface superstructure on the 2DEL at the anatase $TiO_2$ (001)-(1 × 4) surface. a**, *In-situ* LEED pattern of the anatase $TiO_2$ (001) thin film surface taken at 72 V and room temperature showing a two-domain (1 × 4) superstructure. **b**, ARPES Fermi surfaces measured using circularly polarized light at $hv$ = 47 eV and $T$ = 16 K integrated within $E_F$±25 meV. **c**, Schematic Fermi surfaces of the $n$ = 1, 2 subbands. The perpendicular red and black contours arise from the two-domain (1 × 4)

superstructure. Note that the $n = 2$ subband is suppressed in our data from the first Brillouin zone but clearly discernible at $k_x = 2\pi/a$. **d**, Raw photoemission intensity and corresponding curvature plot along the dashed line in **b**. The folded bands and the three subbands are clearly observed. The inset in **d** shows the band folding and the opening of a super-lattice gap at the reduced zone boundary. The data were measured at $hv = 43$ eV. **e**, Comparison of EDCs taken at the points marked by red and blue crosses in panel **b** and the inset of panel **e**. **f**, Lorentzian fits of the peak position, showing a peak shift of ~ 20 meV.

Figure 4. **Calculations of the surface superstructure effect on the 2DEL. a**, Structural model of the anatase TiO$_2$ (001)-(1 × 4) reconstruction proposed by Lazzeri and Selloni[22,23] with the surface Ti atoms labeled by Roman numerals. The top panel shows the lattice potential modulations induced by the (1 × 4) reconstruction along the $x$ direction determined from the surface site-resolved Ti 3$s$ core-level spectra in **b** with respect to the central Ti III atom. **c**, Ti 3$d$ density of states of the surface (black curve) and subsurface (red curve) layer, respectively. The inset shows that the density of states near the bottom of the conduction band arises mostly from subsurface Ti atoms, indicating that the 2DEL resides in the subsurface layers, similar to the 2DEL at the SrTiO$_3$ (110) surface[3]. **d**, Spectral weight distribution in our *ab initio* tight-binding calculations of the surface electronic structure (for details see main text and Methods). **e**, Magnified band dispersion along the white lines in **d** showing the opening of a unidirectional band gap. **f**, Surface density of states (DOS) obtained by weighting the DOS of the first three subbands in our tight-binding calculations with the amplitude of the wave function in the first unit cell containing two TiO$_2$ planes. Higher subbands have negligible in the first unit cell and are not included in the calculations.

1515

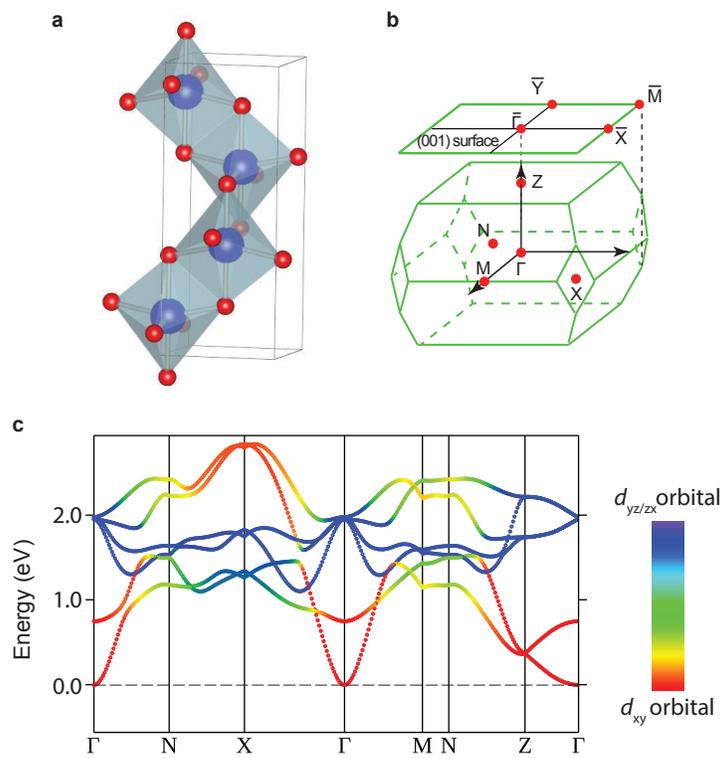

Figure 1

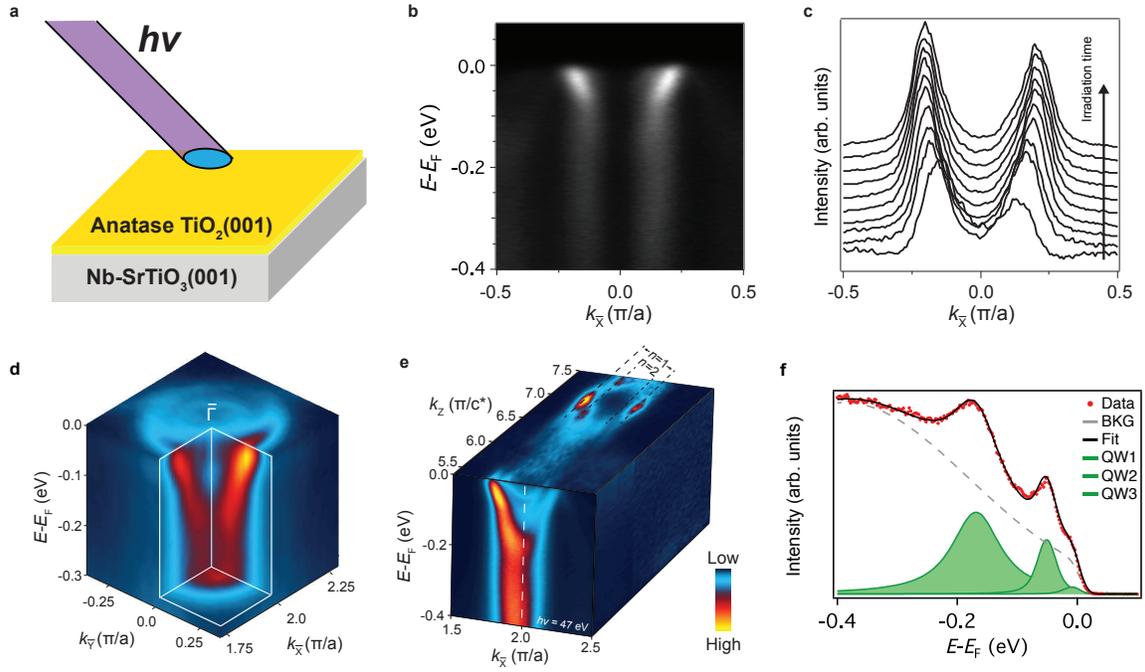

Figure 2

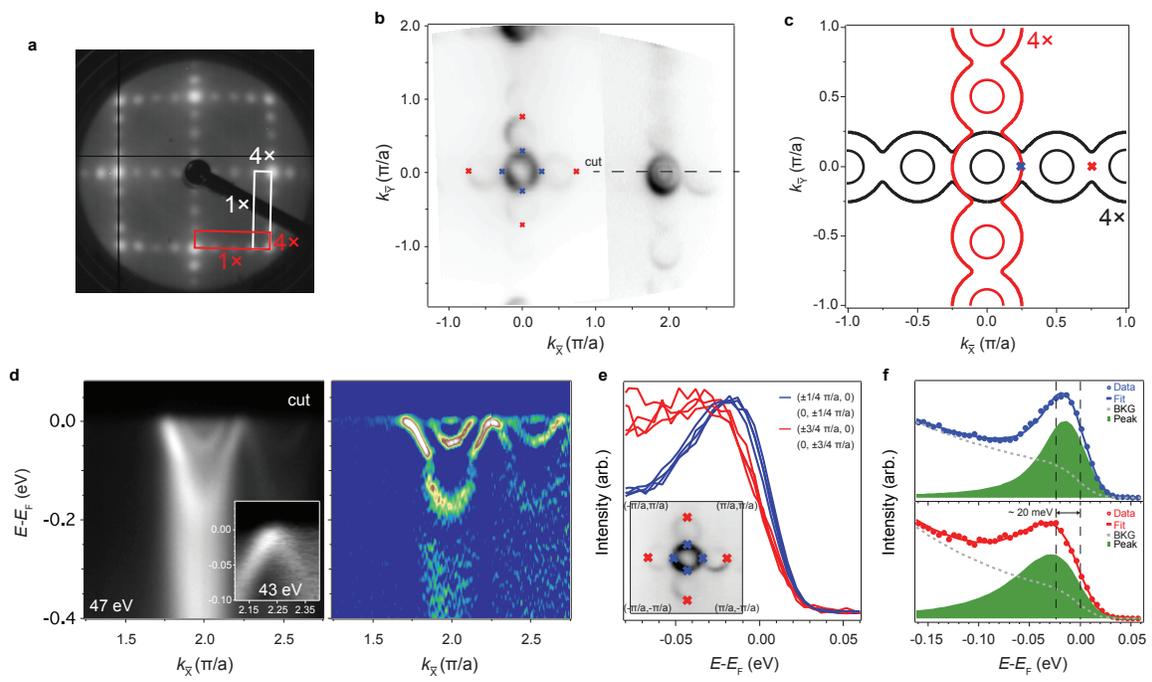

Figure 3

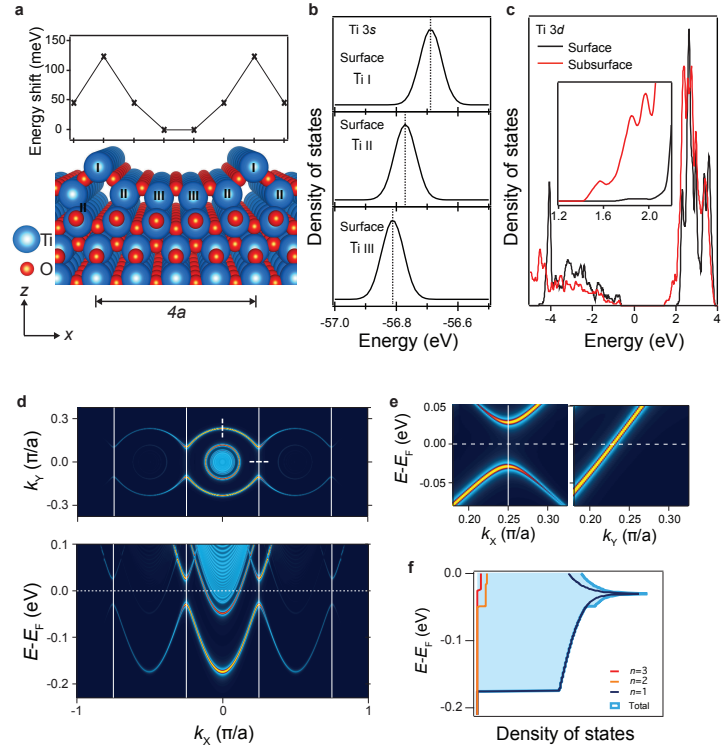

Figure 4